\documentclass[a4paper,11pt,dvitex]{article}
\usepackage{pos}

\newcommand{\be}{\begin{equation}}
\newcommand{\ee}{\end{equation}}
\newcommand{\bea}{\begin{eqnarray}}
\newcommand{\eea}{\end{eqnarray}}
\newcommand{\bi}{\begin{itemize}}
\newcommand{\ei}{\end{itemize}}
\newcommand{\ben}{\begin{enumerate}}
\newcommand{\een}{\end{enumerate}}
\newcommand{\bt}{\begin{tabbing}}
\newcommand{\et}{\end{tabbing}}
\newcommand{\nn}{\nonumber}

\newcommand{\pp}{{p^\prime}}
\newcommand{\bfp}{{\bf p}}
\newcommand{\bfpp}{{{\bf p}^\prime}}

\newcommand{\bfx}{{\bf x}}
\newcommand{\bfxp}{{{\bf x}^\prime}}

\newcommand{\bfz}{{\bf 0}}
\newcommand{\dt}{{\Delta t}}
\newcommand{\dtp}{{\Delta t^\prime}}

\newcommand{\bfxsrc}{{{\bf x}_{\rm src}}}
\newcommand{\vp}{v^\prime}

\newcommand{\bfppp}{{\bf p}_\perp^\prime}

\newcommand{\epsc}{{\epsilon_c:}}
\newcommand{\epsb}{{\epsilon_b}}
\newcommand{\xipi}{{\xi_\pi}}
\newcommand{\xietas}{{\xi_{\eta_s}}}
\newcommand{\xia}{{\xi_a}}
\newcommand{\xiamb}{{\xi_{am_b}}}
\newcommand{\gPVpi}{{g_{D^*D\pi}}}

\title{
   \begin{picture}(0,0)(0,0)%
   \put(350,75){\makebox(0,0)[l]{\textnormal{\normalsize KEK-CP-387}}}%
   \end{picture}
   $B\!\to\!D^{(*)}\ell\nu$ semileptonic decays in lattice QCD
   with domain-wall heavy quarks
}   

\author*[a,b]{T. Kaneko}
\author[c]{Y. Aoki}
\author[d]{B. Colquhoun}
\author[e]{M. Faur}
\author[f]{H. Fukaya}
\author[a,b]{S. Hashimoto}
\author[g]{J. Koponen}
\author[h]{E. Kou}

\affiliation[a]{
High Energy Accelerator Research Organization (KEK),
Ibaraki 305-0801, Japan
}

\affiliation[b]{
School of High Energy Accelerator Science,
SOKENDAI (The Graduate University for Advanced Studies),
Ibaraki 305-0801, Japan
}

\affiliation[c]{
RIKEN Center for Computational Science,
Kobe 650-0047, Japan
}

\affiliation[d]{
Department of Physics and Astronomy,
York University, Toronto, Ontario,
M3J 1P3, Canada
}

\affiliation[e]{
D\'epartement de Physique de l'Ecole
Normale Sup\'erieure - PSL, 75005 Paris, France  
}

\affiliation[f]{
Department of Physics, Osaka University, 
Osaka 560-0043, Japan
}

\affiliation[g]{
PRISMA+ Cluster of Excellence \& Institute f\"ur Kernphysik,
Johannes Gutenberg-Universit\"at Mainz,
D-55128 Mainz, Germany
}

\affiliation[h]{
Universit\'e Paris-Saclay CNRS/IN2P3, IJCLab, 91405 Orsay, France
}

\emailAdd{takashi.kaneko@kek.jp}

\abstract{
We calculate the $B\!\to\!D^{(*)}\ell\nu$ form factors in 2+1 flavor 
relativistic lattice QCD by employing the M\"obius domain-wall action
for all quark flavors.
Our simulations are carried out at lattice cut-offs $a^{-1} \sim 2.5$,
3.6 and 4.5 GeV with the bottom quark masses up to 0.7 $a^{-1}$
to control discretization effects.
We extrapolate the form factors to the continuum limit and
physical quark masses, and discuss systematic uncertainties
of the form factors.
} 

\FullConference{%
 The 38th International Symposium on Lattice Field Theory, LATTICE2021
  26th-30th July, 2021
  Zoom/Gather@Massachusetts Institute of Technology
}

\begin{document}
\maketitle


\section{Introduction}


The $B \to D^{(*)}\ell\nu$ semileptonic decays provide a determination of
the Cabibbo-Kobayashi-Maskawa (CKM) matrix element $|V_{cb}|$,
and are promising probes of new physics, such as the charged Higgs bosons
predicted by supersymmetry. 
However, the current estimate of $|V_{cb}|$ shows a tension
with its alternative determination from the inclusive decay~\cite{HFLAV}.
It is unlikely that the tension is a sign of new physics~\cite{Vcb:NP}
and, hence, we need deeper understanding and better control of
theoretical and experimental uncertainties.
In this article,
we report on our calculation of the $B\!\to\!D^{(*)}\ell\nu$ form factors,
which are the source of the largest theoretical uncertainty of $|V_{cb}|$.


\section{Simulation method}

\begin{table}[b]
\centering
\small
\caption{
  Simulation parameters. Light and strange quark masses, $m_{ud}$ and $m_s$,
  represent their bare values in lattice units.
}
\label{tbl:sim:param}
\begin{tabular}{l|llll}
   \hline 
   lattice parameters 
   & $m_{ud}$ & $m_s$ & $M_\pi$[MeV] & $M_K$[MeV] 
   \\ \hline
   $\beta\!=\!4.17$, \ 
   $a^{-1}\!=\!2.453(4)$, \ 
   $32^3\!\times\!64\!\times\!12$
   & 0.0190 & 0.0400 & 499(1) & 618(1) 
   \\
   & 0.0120 & 0.0400 & 399(1) & 577(1) 
   \\
   & 0.0070 & 0.0400 & 309(1) & 547(1) 
   \\
   & 0.0035 & 0.0400 & 230(1) & 527(1) 
   \\ \hline
   $\beta\!=\!4.17$, \ 
   $48^3\!\times\!96\!\times\!12$
   & 0.0035 & 0.0400 & 226(1) & 525(1) 
   \\ \hline
   $\beta\!=\!4.35$, \ 
   $a^{-1}\!=\!3.610(9)$, \ 
   $48^3\!\times\!96\!\times\!8$
   & 0.0120 & 0.0250 & 501(2) & 620(2) 
   \\
   & 0.0080 & 0.0250 & 408(2) & 582(2) 
   \\
   & 0.0042 & 0.0250 & 300(1) & 547(2) 
   \\ \hline
   $\beta\!=\!4.47$, \ 
   $a^{-1}\!=\!4.496(9)$, \ 
   $64^3\!\times\!128\!\times\!8$
   & 0.0030 & 0.0150 & 284(1) & 486(1) 
   \\ \hline
\end{tabular}
\end{table}

We simulate $N_f\!=\!2+1$ lattice QCD
at cutoffs of $a^{-1}\!\simeq\!2.5$\,--\,4.5~GeV.
The pion mass is as low as $M_\pi\!\sim\!230$~MeV,
and the strange quark mass is close to its physical value.
The M\"obius domain-wall action~\cite{MDWF,MDWF:JLQCD:lat13}
is employed for all relevant quark flavors
to preserve chiral symmetry to good accuracy,
which simplifies the renormalization of the relevant matrix elements.
The charm quark mass $m_c$ is set to its physical value fixed from
the spin averaged mass $(M_{\eta_c}+3M_{J/\Psi})/4$.
Depending on the lattice spacing $a$, 
we take three to six values of the bottom quark mass
$m_b\!=\!1.25^n m_c (n=0,1,...)$ below $0.7\,a^{-1}$ 
in order to control $O((am_b)^2)$ discretization effects.
The spatial lattice size $L$ is chosen to satisfy
the condition $M_\pi L \!\gtrsim\!4$ to suppress finite volume effects (FVEs).
At our smallest $M_\pi\!\simeq\!230$~MeV,
we simulate a smaller lattice size $M_\pi L \!\sim\!3$
to directly study the FVEs.
The statistics are 5,000 Molecular Dynamics time
at each simulation point.
These simulation parameters are listed in Table~\ref{tbl:sim:param}.
The same gauge ensembles are used in our study of
the $B\!\to\!\pi\ell\nu$~\cite{Nf2+1:B2pi:JLQCD:N}
and inclusive decays~\cite{incl}
and in a joint project with the RBC/UKQCD Collaboration
on the $B$ meson mixing~\cite{mixing}.


The $B\!\to\!D^{(*)}$ matrix elements are parametrized
by six form factors in total,
\bea
   \sqrt{M_B M_D}^{-1}
   \langle D(\pp) | V_\mu | B(p) \rangle
   & = &
   (v+\vp)_\mu h_+(w) + (v-\vp)_\mu h_-(w),
   \label{eqn:ff:B2D}
   \\[2mm]
   \sqrt{ M_B M_{D^*} }^{-1}
   \langle D^*(\epsilon,\pp) | V_\mu | B(p) \rangle
   & = &
   i \varepsilon_{\mu\nu\rho\sigma} \, \epsilon^{*\nu} v^{\prime\rho} v^\sigma \, h_V(w),
   \label{eqn:ff:B2D*:V}
   \\[2mm]
   \sqrt{ M_B M_{D^*} }^{-1}
   \langle D^*(\epsilon,\pp) | A_\mu | B(p) \rangle
   & = &
   (w+1) \, \epsilon_\mu^* \,   h_{A_1}(w)
   - (\epsilon^* v)\, v_\mu \, h_{A_2}(w) - (\epsilon^* v)\, \vp_\mu \, h_{A_3}(w),
   \hspace{10mm}
   \label{eqn:ff:B2D*:A}
\eea
where
$w=v\vp$ is the recoil parameter
defined by four velocities $v\!=\!p/M_B$ and $\vp\!=\!\pp/M_{D^{(*)}}$,
and $\epsilon$ is the polarization vector of $D^*$,
which satisfies $\pp\epsilon\!=\!0$.


The $B\!\to\!D^{(*)}$ matrix elements can be extracted from 
three- and two-point functions, provided that they are dominated by
their ground state contribution as 
\bea
   &&
   C_{\mathcal O_\Gamma}^{BD^{(*)}}(\dt,\dtp;\bfp,\bfpp)
   \nn \\
   & = & 
   \sum_{\bfxsrc,\bfx,\bfxp}
   \langle 
      {\mathcal O}_{D^{(*)}}(\bfxp,\dt+\dtp)
      {\mathcal O}_\Gamma(\bfx,\dt)
      {\mathcal O}_{B}(\bfxsrc,0)^\dagger
   \rangle
   e^{-i\bfp(\bfx-\bfxsrc)-i\bfpp(\bfxp-\bfx)}
   \nn \\
   & \xrightarrow[\dt,\dtp \to \infty ]{} &
   \frac{Z_{D^{(*)}}^*(\bfpp)\,Z_B(\bfp)}{4E_{D^{(*)}}(\bfpp)E_B(\bfp)}
   \langle D^{(*)}(\pp) | {\mathcal O}_\Gamma | B(p) \rangle
   e^{-E_{D^{(*)}}(\bfpp)\dtp -E_B(\bfp)\dt }.
   \label{eqn:ff:corr_3pt}
\eea
Here ${\mathcal O}_\Gamma\!=\!V_\mu$ or $A_\mu$,
and the argument $\epsilon$ is suppressed for $Z_{D^*}$ and $|D^*(\pp)\rangle$
for simplicity.
We apply Gaussian smearing to the interpolating fields
${\mathcal O}_P$ of the mesons $P\!=\!B,D,D^*$
to enhance their overlap with  the ground state
$Z_P(\bfp)\!=\!\langle P(p) | {\mathcal O}_P^\dagger \rangle$.
The $B$ meson is at rest ($\bfp\!=\!\bfz$) throughout our measurements,
whereas we vary the three momentum of $D^{(*)}$ as 
$|\bfpp|^2\!=\!0,1,2,3,4$ in units of $(2\pi/L)^2$
in order to study the $w$ dependence of the form factors.


We construct ratios of the correlation functions as 
\bea
   R_{(k)}^{BD^{(*)}}(\dt,\dtp)
   & = &
   \frac{ C_{V_4(A_k)}^{BD^{(*)}}(\dt,\dtp;\bfz,\bfz)\,
          C_{V_4(A_k)}^{D^{(*)}B}(\dt,\dtp;\bfz,\bfz) }
        { C_{V_4}^{BB}(\dt,\dtp;\bfz,\bfz)\,
          C_{V_4}^{D^{(*)}D^{(*)}}(\dt,\dtp;\bfz,\bfz) }
   \xrightarrow[\dt,\dtp\to\infty]{}
   |h_{+(A_1)}(1)|^2,
   \label{eqn:ff:b2d+b2d*:r1}
   \\[-1mm]
   R_{V,k}^{BD^*}(\dt,\dtp;\bfz,\bfppp)
   & = &
   \frac{ C_{V_k}^{BD^*}(\dt,\dtp;\bfz,\bfppp) }
        { C_{A_k}^{BD^*}(\dt,\dtp;\bfz,\bfppp) }
   \xrightarrow[\dt,\dtp\to\infty]{}
   \frac{ \epsilon_{kij} \epsilon^*_i v_{\perp j}^\prime}{1+w}
   \frac{h_V(w)}{h_A(w)}
   \label{eqn:ff:b2d*:r3}
\eea
where $\bfppp$ denotes the $D^*$ momentum satisfying $v \epsilon\!=\! 0$.
Unnecessary factors, such as the overlap $Z_P(\bfp)$
and exponential damping factors in Eq.~(\ref{eqn:ff:corr_3pt}),
cancel in the ratios~\cite{double_ratio}.
We also expect a partial cancellation of statistical fluctuations.
These ratios are used to precisely extract the form factors at zero recoil,
$h_+(1)$ and $h_{A_1}(1)$ and a ratio $R_1(w)\!=\!h_V(w)/h_{A_1}(w)$,
which are key hadronic inputs in the conventional determination of $|V_{cb}|$.
%
We refer the reader to Refs.~\cite{B2D*:Nf2+1:JLQCD:lat18,B2D*:Nf2+1:Fermilab/MILC:wne1} for ratios to extract other form factors.

\begin{figure}[tb] 
  \centering
  \includegraphics[angle=0,width=0.6\linewidth,clip]%
                  {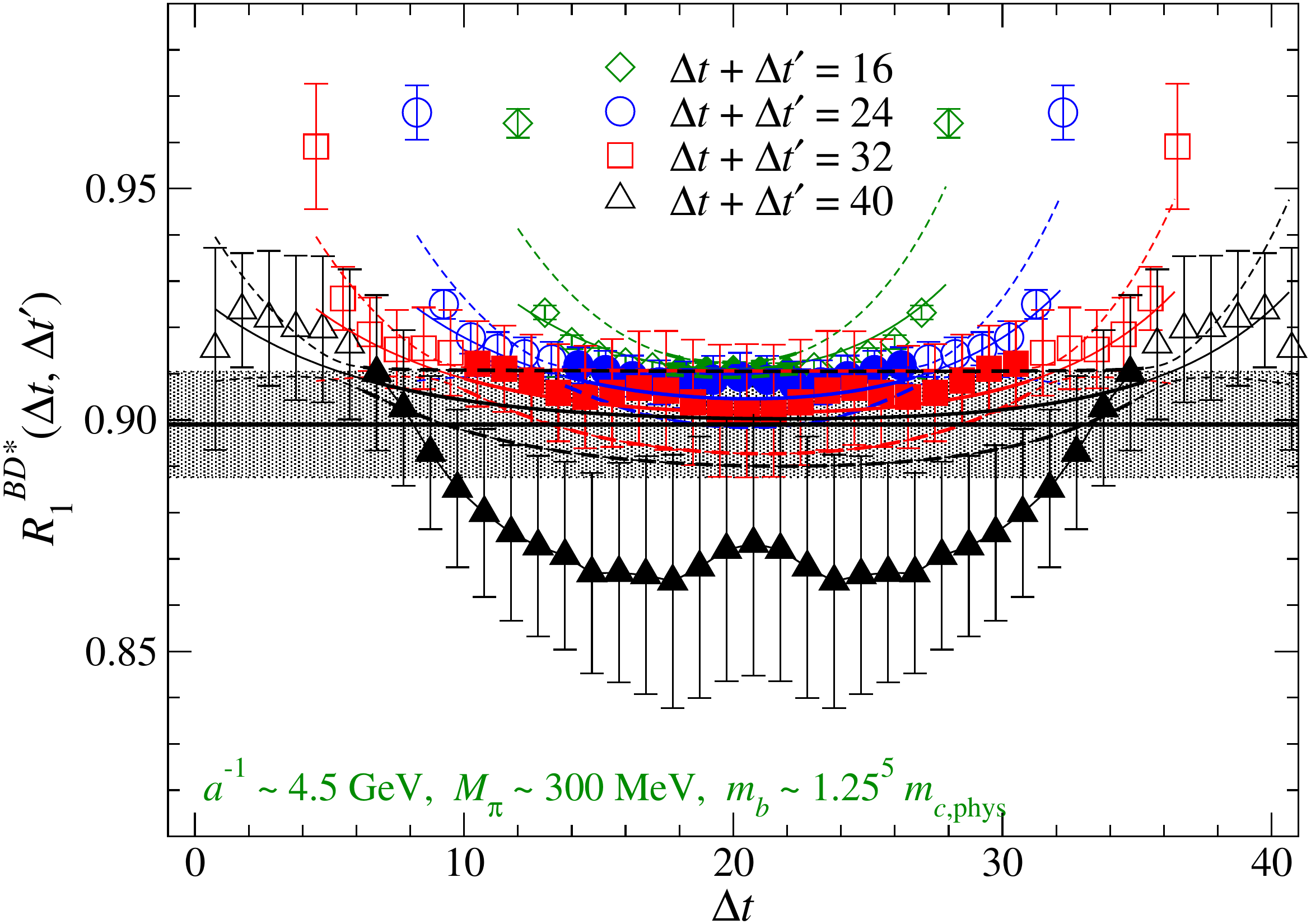}
  \caption{
    Double ratio $R_1^{BD^{*}}(\dt,\dtp)$
    as a function of the temporal location $\Delta t$ of $A_1$.
    We plot data at $\beta\!=\!4.35$, $m_{ud}\!=\!0.0042$
    and $m_b\!=\!1.25^5\,m_c$.
    Diamonds, circles, squares and triangles show data
    with the source\,--\,sink separation
    $\dt+\dtp\!=\!16$, 24, 32 and 40, respectively.
    We also plot the multi-exponential fit~(\ref{eqn:ff:b2d*:r1:fit})
    by the dashed lines,
    and the ground state contribution $|h_{A_1}(1)|^2$
    by the black shaded band.
  }
  \vspace{0mm}
  \label{fig:ff:drat14}
\end{figure}

In order to control the excited state contamination,
we calculate the correlator ratios
with four different values of the source-sink separation $\dt+\dtp$,
and then extract form factors by using multi-exponential fits.
The fitting form, for instance, to determine $h_{A_1}(1)$ is 
\bea
   R_1^{BD^{*}}(\dt,\dtp)
   & = &
   |h_{A_1}(1)|^2
   \left(
      1 + a e^{-\Delta M_B \dt}  + b e^{-\Delta M_{D^*} \dt}
        + c e^{-\Delta M_B \dtp} + d e^{-\Delta M_{D^*} \dtp}
   \right),
   \label{eqn:ff:b2d*:r1:fit}
   \eea
where $\Delta M_{B(D^*)}$ represents the energy difference
between the $B(D^*)$ meson ground state 
and the first excited state with the same quantum numbers,
and is estimated from two-point functions of $B$ ($D^*$). 
As plotted in Fig.~\ref{fig:ff:drat14},
the multiple values of $\dt+\dtp$ enable us to safely identify
the ground state contribution,
whereas data at small $\dt+\dtp$ are helpful in achieving 
good statistical accuracy.

A salient feature of our simulations with relativistic heavy quarks
and chiral symmetry 
is that the form factors in the Standard Model,
namely those in Eqs.~(\ref{eqn:ff:B2D})\,--\,(\ref{eqn:ff:B2D*:A}),
can be calculated without finite renormalization of
the lattice operators $V_\mu$ and $A_\mu$.
Our correlator ratios are constructed so that
renormalization factors cancel as seen
in Eqs.~(\ref{eqn:ff:b2d+b2d*:r1}) and (\ref{eqn:ff:b2d*:r3}).
This is an important advantage,
because it is not straightforward to reliably estimate
and reduce higher order corrections in the perturbative renormalization
and matching.


\begin{figure}[tb] 
  \centering
  \includegraphics[width=0.49\linewidth,clip]{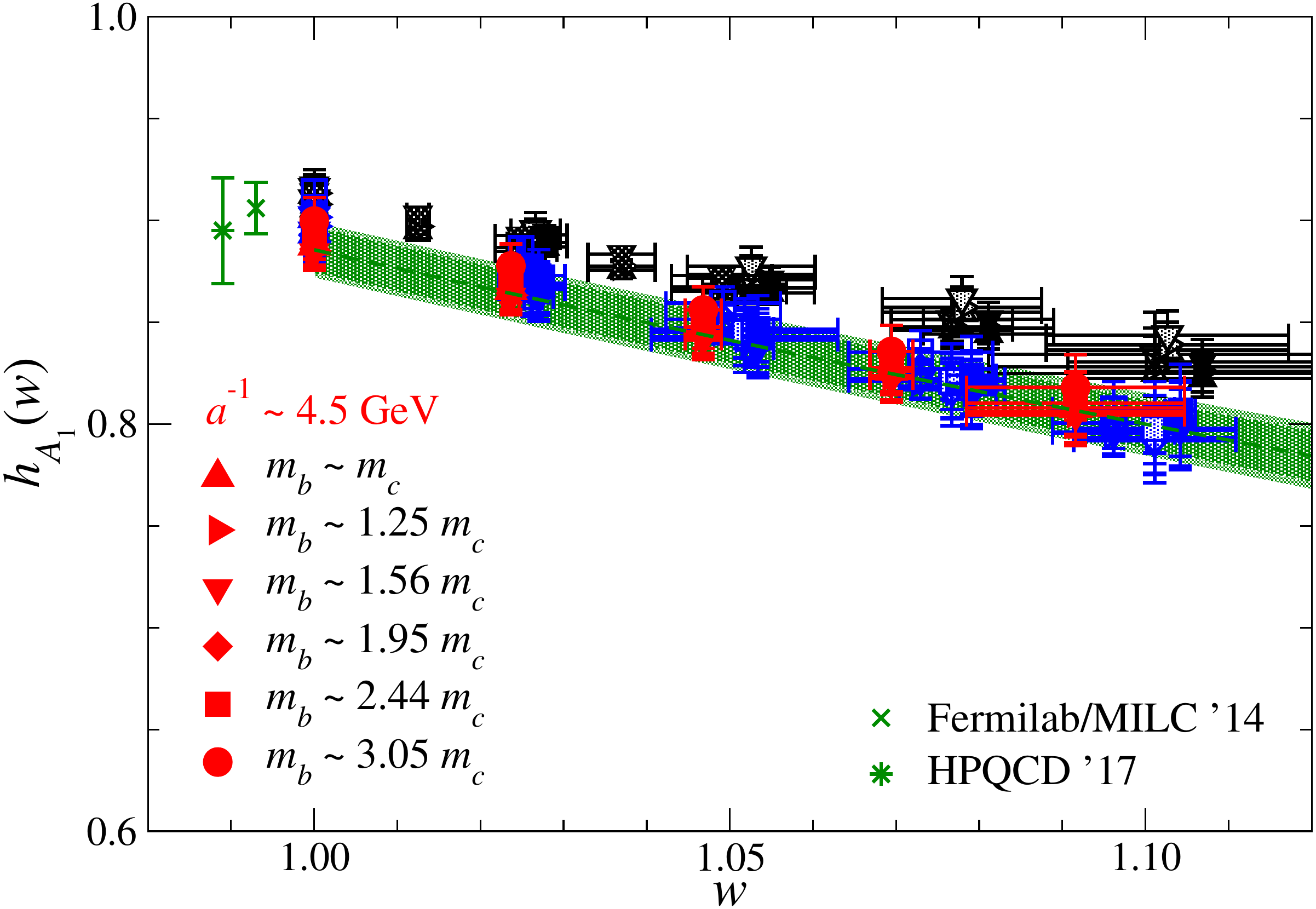}
  \hspace{1mm}
  \includegraphics[width=0.49\linewidth,clip]{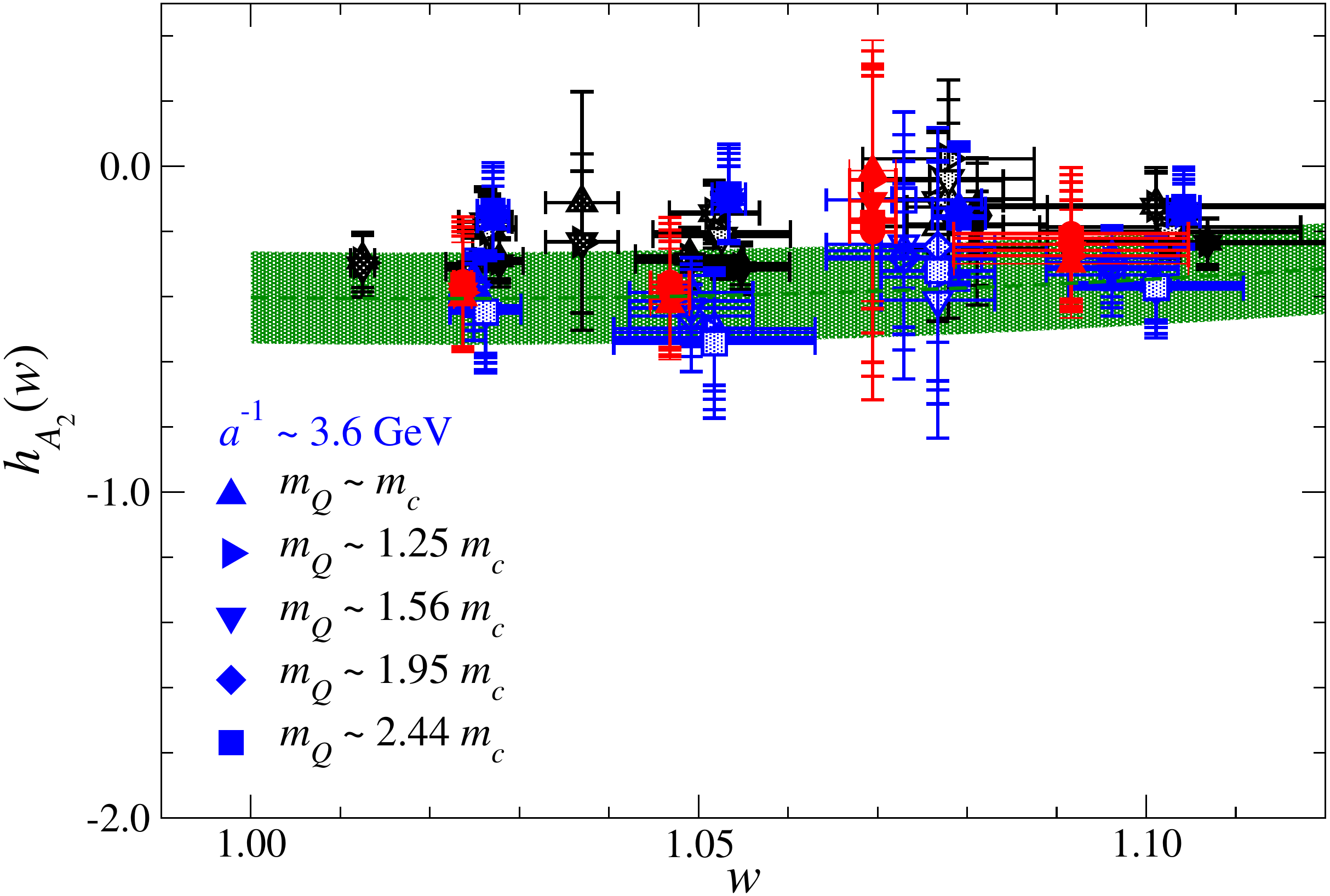}
  \vspace{-3mm}
  
  \includegraphics[width=0.49\linewidth,clip]{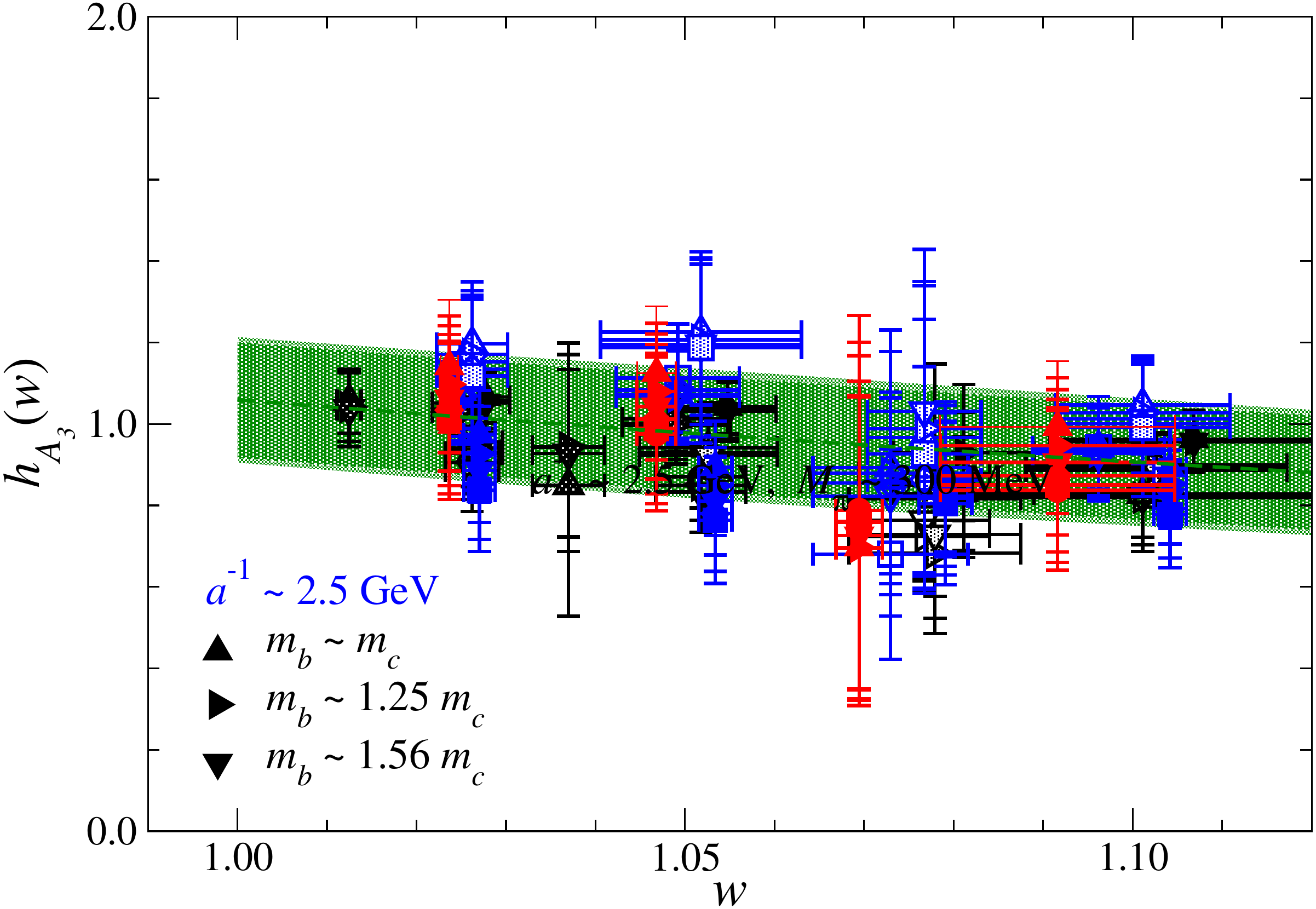}
  \hspace{1mm}
  \includegraphics[width=0.49\linewidth,clip]{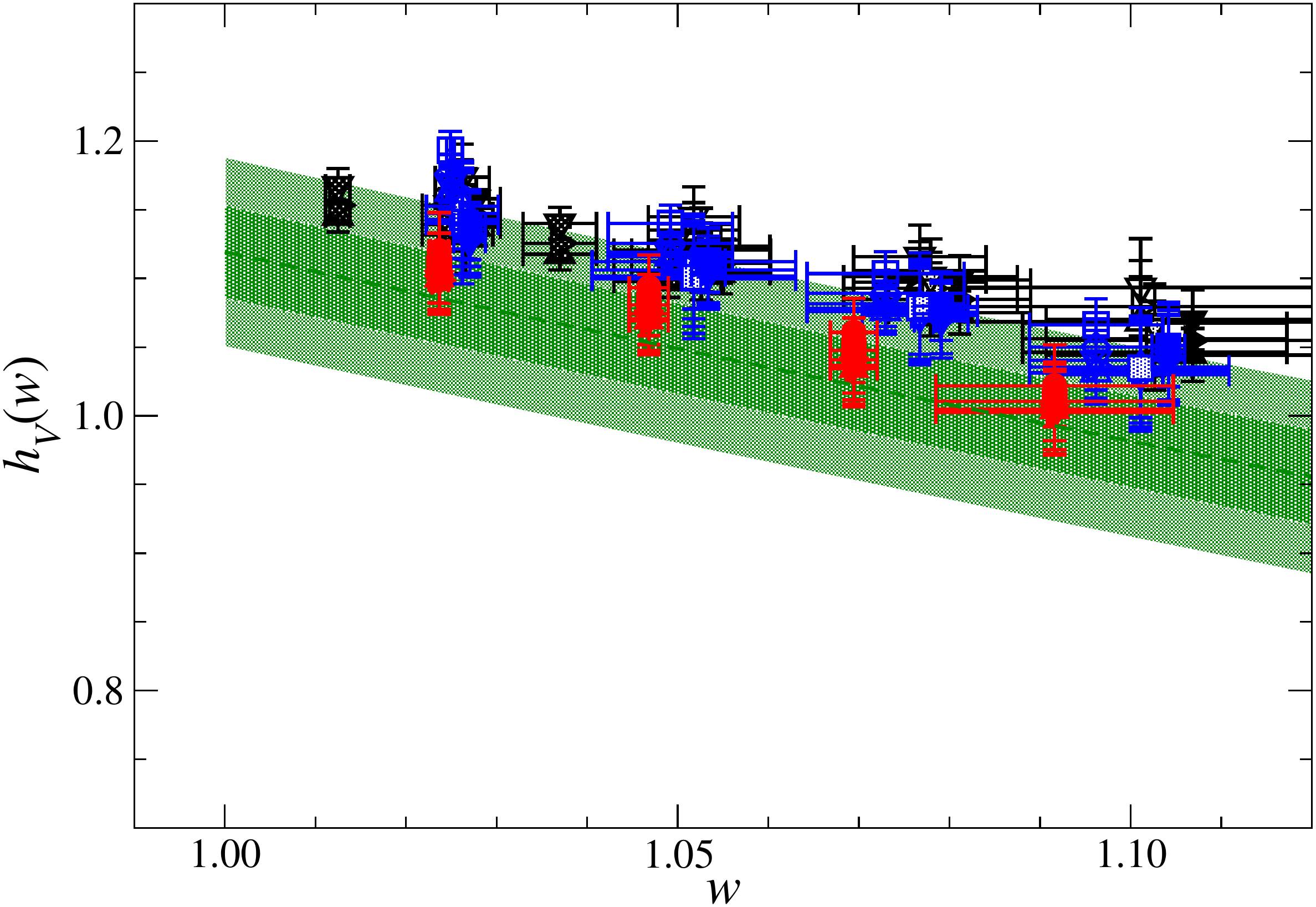}
  \vspace{-3mm}

  \caption{
    $B\!\to\!D^*\ell\nu$ form factors,
    $h_{A_1}$ (top-left), $h_{A_2}$ (top-right), $h_{A_3}$ (bottom-left)
    and $h_V$ (bottom-right) as a function of $w$.
    In all panels,
    we plot data at $a^{-1}\!=\!2.5$, 3.6 and 4.5~GeV
    by the black, blue and red symbols, respectively.
    The open, pale shaded, filled, dark shaded symbols show
    data at $M_\pi\!\sim\!500$, 400, 300 and 230~MeV, whereas
    symbols with different shapes are data with different values of $m_b$.
    We plot the form factors extrapolated to the continuum limit
    and physical quark masses by the green bands,
    and their statistical (total) uncertainty is shown by
    the dark (pale) bands.
    We also plot the previous estimates of
    $h_{A_1}(1)$~\cite{B2D*:Nf2+1:Fermilab/MILC:w1,B2D*:HPQCD:w1},
    which is shifted in the horizontal direction for clarity.
  }
  \label{fig:ccfit:B2D*}
  \vspace{-3mm}
\end{figure}

\section{Continuum and chiral extrapolation}


We extrapolate the form factors to the continuum limit
and physical quark masses by employing a fitting form
with a chiral logarithm predicted by 
next-to-leading order 
heavy meson chiral perturbation theory~\cite{B2D*:HMChPT:RW,B2D*:HMChPT:S}
and polynomial corrections.
The fitting form for $h_{A_1}$, for instance, is 
\bea
   \frac{h_{A_1}}{\eta_A}
   & = &
   c
 + c_w (w-1) + d_w (w-1)^2
 + c_b (w-1) \epsb + d_b \epsb^2
 \nn \\
   & &
 + \frac{\gPVpi^2}{32 \pi^2 f_\pi^2}
   \bar{F}\!\left(\xipi,\Delta_c\right)
 + c_\pi \xipi + c_{\eta_s} \xietas
 + c_a \xia + c_{am_b} \xiamb,
 \label{eqn:chiral_fit:form:w-rad}
\eea
where $\eta_A$ represents the one-loop radiative correction
evaluated by using our estimate of the strong coupling
from charmonium correlators~\cite{alphas:Nf2+1:JLQCD}.
The polynomial corrections are written
in terms of small expansion parameters
\bea
   w-1,
   \hspace{2mm}
   \epsb   = \frac{\bar{\Lambda}}{2m_b},
   \hspace{2mm}
   \xipi   = \frac{M_\pi^2}{(4 \pi f_\pi)^2},
   \hspace{2mm}
   \xietas = \frac{M_{\eta_s}^2}{(4 \pi f_\pi)^2},
   \hspace{2mm}
   \xia    = (\Lambda a)^2,
   \hspace{2mm}
   \xiamb  = (am_b)^2,
   \label{eqn:chiral_fit:expand_param}
\eea
where we set $\bar{\Lambda}, \Lambda\!=\!0.5$~GeV.
For $h_{A_1}$ and $h_+$,
the $O(\epsb)$ term has an additional factor of $w-1$
to be consistent with Luke's theorem~\cite{luke_theorem},
and we include an $O(\epsb^2)$ correction
as in Eq.~(\ref{eqn:chiral_fit:form:w-rad}).


For the chiral expansion,
we employ the so-called $\xi$-expansion in $\xipi$ and $\xietas$,
with which we observe better convergence
for the light meson observables than the $x$-expansion
in $M_\pi^2/(4\pi F)^2$ and $M_{\eta_s}^2/(4\pi F)^2$,
where $F$ is the decay constant in the chiral limit~\cite{chiral_exp:Nf2:JLQCD}.


The explicit form of the loop integral function for the chiral logarithm
is given in Refs.~\cite{B2D*:HMChPT:RW,B2D*:HMChPT:S},
and can be approximated as 
$\bar{F}\!=\!\Delta_c^2 \ln[ \xipi ] + O(\Delta_c^3)$,
where $\Delta_c$ represents the $D^*$\,--\,$D$ mass splitting.
The $D^*D\pi$ coupling is set to $\gPVpi\!=\!0.53(8)$,
which is quoted in Ref.~\cite{B2D*:Nf2+1:Fermilab/MILC:w1}
and the error covers previous estimates.
This choice, however, has small impact
on the continuum and chiral extrapolation,
since the chiral logarithm is suppressed
by the small mass splitting $\Delta_c$.


In Fig.~\ref{fig:ccfit:B2D*},
we plot the $B\!\to\!D^*\ell\nu$ form factors at simulation points
and those extrapolated to the continuum limit and physical quark masses.
With our choice of the expansion parameters (\ref{eqn:chiral_fit:expand_param}),
all coefficients $c_{\{w,b,\pi,\eta_s,a,am_b\}}$ and $d_{\{w,b\}}$ turn out
to be $O(1)$ or smaller. 
The figure suggests that the form factors mildly depend on $a$ and quark masses.
Consequently,
the fitting form (\ref{eqn:chiral_fit:form:w-rad}) 
describes our data well 
with $\chi^2/{\rm d.o.f}\!\sim\!0.2$,
and many coefficients are consistent with zero.
We also note that our result for $h_{A_1}(1)$ is in reasonable agreement with
the previous estimates by Fermilab/MILC~\cite{B2D*:Nf2+1:Fermilab/MILC:w1}
and HPQCD~\cite{B2D*:HPQCD:w1}.


\begin{figure}[tb] 
  \centering
  \includegraphics[angle=0,width=0.6\linewidth,clip]%
                  {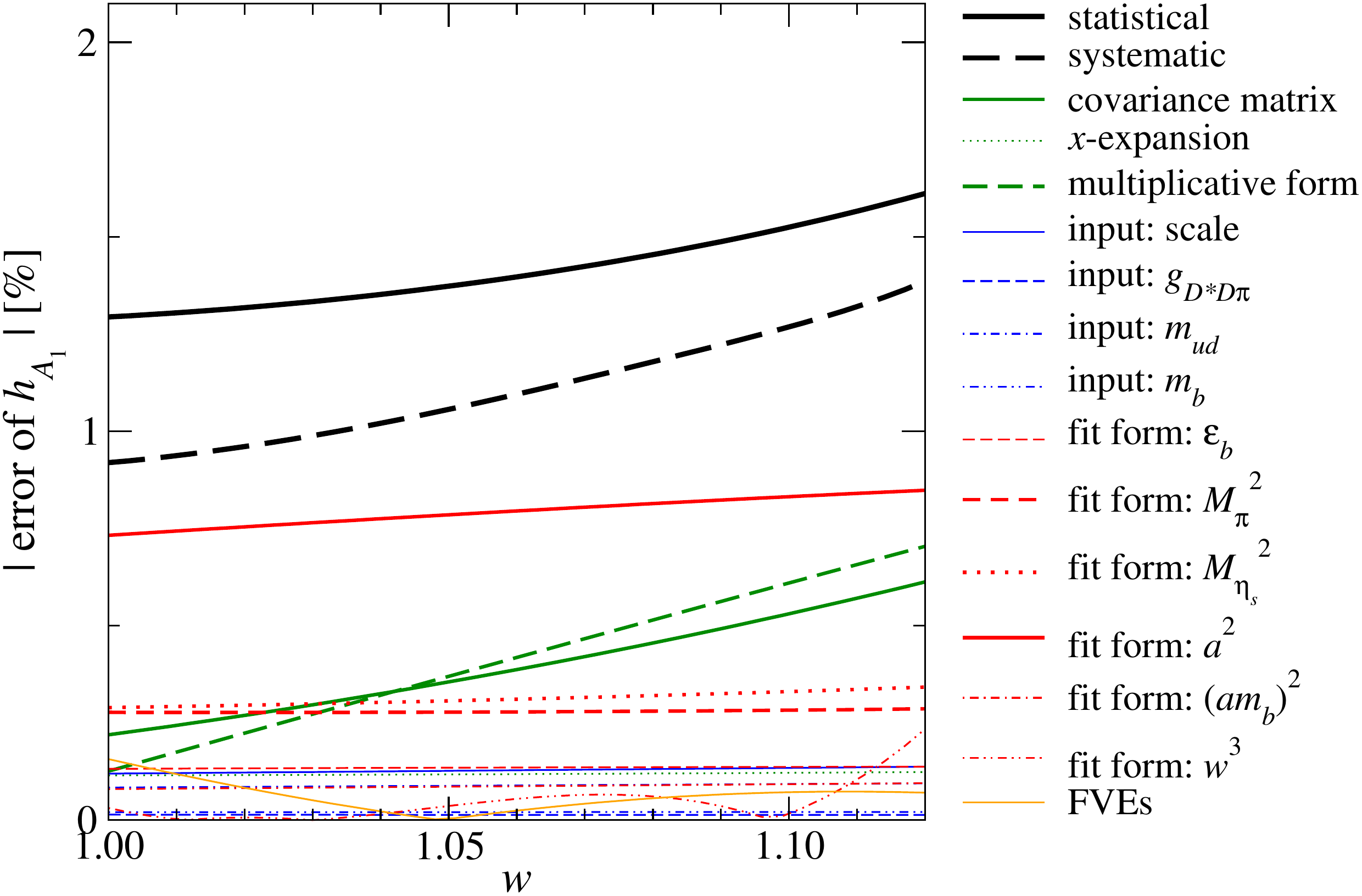}
  \caption{
    Uncertainties of $h_{A_1}(w)$
    in the continuum limit and at physical quark masses.
    We plot the statistical (black thick solid line)
    and total systematic (black thick dashed line) errors,
    and the breakdown of the latter (other thin lines)
    as a function of $w$.
  }
  \vspace{0mm}
  \label{fig:ff:chiral_fit:errors}
\end{figure}

We estimate the following systematic errors, and plot them
in Fig.~\ref{fig:ff:chiral_fit:errors}.
\bi

\item {\bf Covariance matrix} :
With our statistics, the covariance matrix has exceptionally small eigenvalues.
We test the singular value decomposition cut as well as 
the shrinkage method~\cite{shrinkage},
and the change of the form factors is taken as a systematic uncertainty.

\item {\bf Chiral expansion} :
We test the $x$-expansion instead of the $\xi$-expansion.

\item {\bf Multiplicative form} :
We test the following multiplicative fitting form
in order to take account of possible cross terms,
\bea
   \frac{h_{A_1}}{\eta_A}
   & = &
   c - 1 
   +
   \left( 1 + c_w (w-1) + d_w (w-1)^2 \right)
   \left( 1 + c_b (w-1) \epsb + d_b \epsb^2 \right)
   \nn \\
   & &
   \times
   \left( 1 + \frac{\gPVpi^2}{32 \pi^2 f_\pi^2}\bar{F} + c_\pi \xipi \right)
   \left( 1 + c_{\eta_s} \xietas \right)
   \left( 1 + c_a \xia \right)
   \left( 1 + c_{am_b} \xiamb \right).
\label{eqn:chiral_fit:form:multi}
\eea

\item {\bf Inputs} :
We repeat the fit with $a$ and $\gPVpi$ shifted by their uncertainty.

\item {\bf Strong isospin breaking} : 
We shift the input $M_\pi$ to fix $m_{ud}$ 
by a relative difference $2|(M_{\pi^0}-M_{\pi^\pm})|/(M_{\pi^0}+M_{\pi^\pm})$
to take account of the uncertainty due to strong isospin breaking.
We carry out a similar analysis for $M_{\eta_b}$ to fix $m_b$.

\item {\bf Polynomial correction} : 
In order to estimate the systematic uncertainty due to the choice
of the polynomial correction in Eq.~(\ref{eqn:chiral_fit:form:w-rad}),
we repeat the fit by adding a higher order term or
removing a poorly-determined term.

\item {\bf FVEs} : 
The form factor data at $a^{-1}\!=2.453$~GeV, $M_\pi\!=\!226$~MeV
is replaced by those calculated on a smaller lattice $32^3 \times 64$
to estimate FVEs.

\ei
As seen in Fig.~\ref{fig:ff:chiral_fit:errors},
each systematic error of $h_{A_1}$ is suppressed well below
the statistical error of 1\,--\,2\,\%. 
Other form factors have larger and more dominant statistical errors.
With the exception of $h_V$, where the discretization error is 5\,\%
and the statistical error is 3\,\%,
all other systematic errors for each of the form factors are suppressed
such that they are similar or smaller than their corresponding
statistical uncertainties.

Based on the continuum and chiral extrapolation,
we can generate synthetic data of the form factors.
A model independent parametrization of them and
comparison with the previous study~\cite{B2D*:Nf2+1:Fermilab/MILC:wne1}
are in progress~\cite{CKM2021}.


\section{Form factor beyond the SM}

One of the the so-called B physics anomalies has been reported
for the decay rate ratio to measure the lepton flavor universality violation
$R(D^{(*)})\! = \!\Gamma(B\!\to\!D^{(*)}\tau\nu)/\Gamma(B\!\to\!D^{(*)}\ell\nu)$
($\ell=e,\mu$), which currently poses a 3~$\sigma$ tension
between the SM and experiment.
In order to interpret such a hint based on new physics models,
lattice QCD is expected to provide form factors beyond the SM,
namely those for the (pseudo-)scalar and tensor interactions.

\begin{figure}[tb] 
  \centering
  \includegraphics[angle=0,width=0.6\linewidth,clip]%
                  {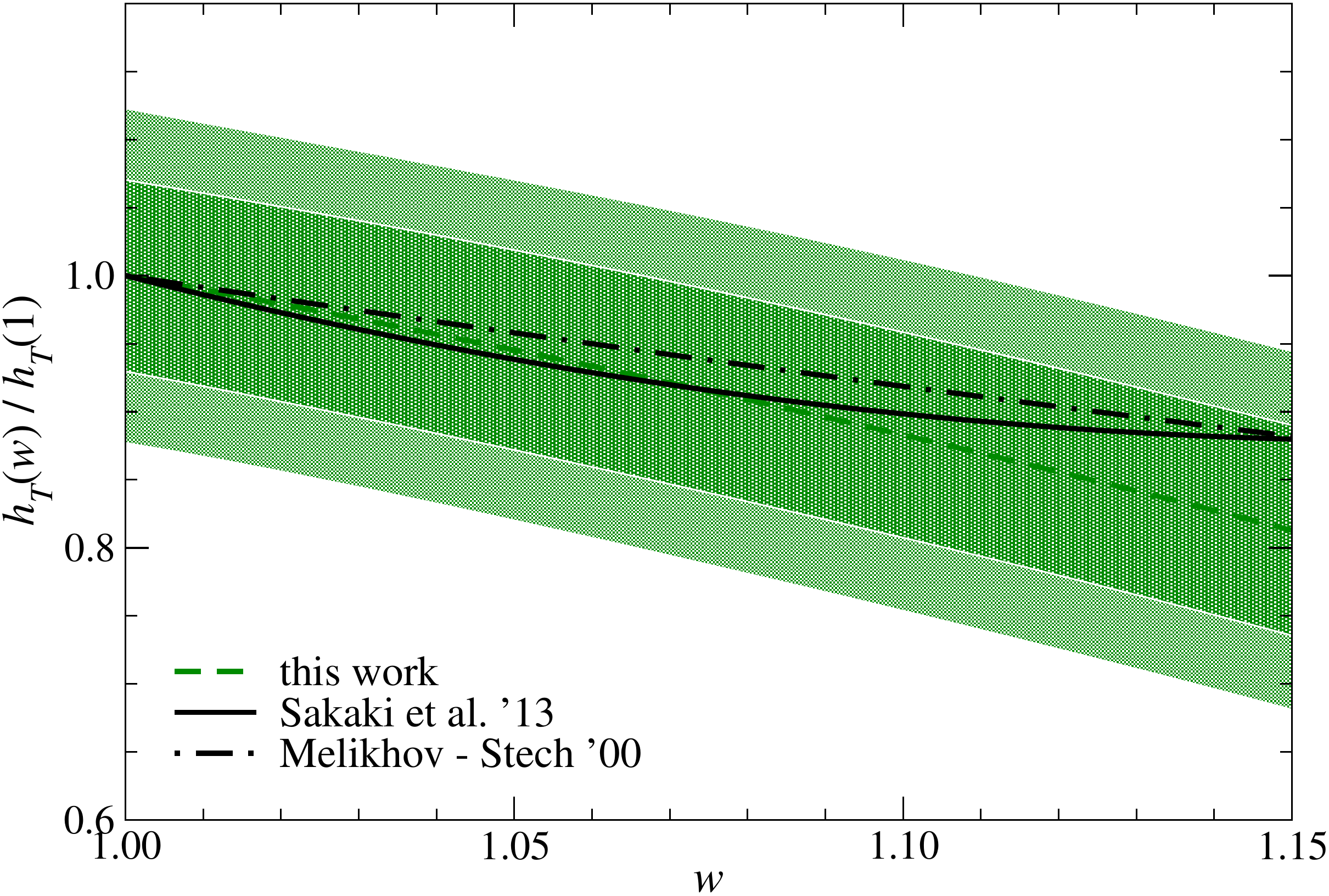}
  \caption{
    Tensor form factor for $B\!\to\!D\ell\nu$ decay
    normalized at $w\!=\!1$.
    Our result extrapolated to the continuum limit and physics quark masses
    is plotted by the green dashed line
    with the dark and pale green bands representing
    the statistical and total uncertainties, respectively.
    We also plot phenomenological estimates in Refs.~\cite{hT:eom,hT:reso}.
  }
  \vspace{-5mm}
  \label{fig:tensor:ff}
\end{figure}

Figure~\ref{fig:tensor:ff} shows our preliminary results for
the $B\!\to\!D\ell\nu$ tensor form factor $h_T$ defined by
\bea
   \sqrt{M_B M_D}^{-1}
   \langle D(\pp) \left| \bar{c}\sigma_{\mu\nu} b \right| B(p) \rangle
   & = &
   i h_T(w) ( \vp_\mu v_\nu - \vp_\nu v_\mu ).
\eea
We use a correlator ratio
\bea
   R_T^{BD}(\dt,\dtp;\bfz,\bfp)
   & = &
   \frac{ C_{T_{k4}}^{BD}(\dt,\dtp;\bfz,\bfp)\,
          C^D(\dtp;\bfz) }
        { C_{V_4}^{BD}(\dt,\dtp;\bfz,\bfz)\,
          C^D(\dtp;\bfp) }
   \xrightarrow[\dt,\dtp\to\infty]{}
   \frac{v_k}{2}\frac{h_T(w)}{h_+(1)},
   \label{eqn:tensor:dble_rat}
\eea
where $C^D$ represents the $D$ meson two-point function,
and $h_+(1)$ is calculated from Eq.~(\ref{eqn:ff:b2d+b2d*:r1}).
In contrast to the calculation of the SM form factors,
renormalization factors do not cancel in the ratio $R_T^{BD}$.
A non-perturbative renormalization based on the Borel transformation
of the hadronic vacuum polarization function is in progress~\cite{NPR}.
In the following, we consider 
$h_T(w)$ normalized at $w\!=\!1$ so that renormalization factor cancels.

We observe that the tensor form factor mildly depends on the lattice spacing
and quark masses, and carry out the continuum and chiral extrapolation
by using a fitting form similar to Eq.~(\ref{eqn:chiral_fit:form:w-rad}).
Figure~\ref{fig:tensor:ff} shows $h_T(w)/h_T(1)$ in the continuum limit
and at physical quark masses. 
The preliminary result has 10\,\% statistical and 10\,\% systematic errors,
and is consistent with previous phenomenological estimates~\cite{hT:eom,hT:reso}.

\section{Summary}

In this article,
we report on our study of the $B\!\to\!D^{(*)}\ell\nu$ form factors
at non-zero recoil in lattice QCD with domain-wall heavy quarks.
The SM form factors do not need explicit renormalization
thanks to chiral symmetry.
The mild dependence of the form factors on simulation parameters
brings the continuum and chiral extrapolation under reasonable control.
As a result,
the axial form factor $h_{A_1}$ is determined with an accuracy of
$\lesssim 2$\,\% in the simulation region of $w$.
Our study is being extended to form factors beyond the SM,
and we are calculating the $B\!\to\!D\ell\nu$ tensor form factor.
Implications of our results on the determination of $|V_{cb}|$
and new physics searches are under investigation.


This work used computational resources of supercomputer
Fugaku provided by the RIKEN Center for Computational Science
and Oakforest-PACS provided by JCAHPC
through the HPCI System Research Projects
(Project ID: hp180132, hp190118 and hp210146)
and Multidisciplinary Cooperative Research Program
in Center for Computational Sciences, University of Tsukuba
(Project ID: xg18i016).
This work is supported in part by JSPS KAKENHI Grant Numbers
18H03710 and 21H01085
and by the Toshiko Yuasa France Japan Particle Physics Laboratory
(TYL-FJPPL project FLAV\_03).


\begin{thebibliography}{99}

\bibitem{HFLAV}
  Y.~Amhis {\it et al.} (Heavy Flavor Averaging Group),
  Eur. Phys. J. C 81 (2021) 226.

\bibitem{Vcb:NP}
  A. Crivellin and S. Pokorski,
  Phys. Rev. Lett. 114 (2015) 011802.

\bibitem{MDWF}
  R.C.~Brower, H.~Neff and K.~Orginos,
  Nucl. Phys. (Proc.Suppl.) 140 (2005) 686.

\bibitem{MDWF:JLQCD:lat13}
  T.~Kaneko {\it et al.} (JLQCD Collaboration),
  PoS (LATTICE 2013) 125.

\bibitem{Nf2+1:B2pi:JLQCD:N}
  B.~Colquhoun {\it et al.} (JLQCD Collaboration),
  PoS (LATTICE2019) 143.

\bibitem{incl}
  S.~M\"achler {\it et al.}, PoS (LATTICE2021) 512 in these proceedings;
  S.~Hashimoto {\it et al.}, PoS (LATTICE2021) 534 in these proceedings.

\bibitem{mixing}
  P.~Boyle {\it et al.} (RBC/UKQCD and JLQCD Collaborations),
  PoS (LATTICE2021) 224 in these proceedings.

\bibitem{double_ratio}
  S.~Hashimoto {\it et al.},
  Phys. Rev. D61 (1999) 014502.

\bibitem{B2D*:Nf2+1:JLQCD:lat18}
  T.~Kaneko {\it et al.} (JLQCD Collaboration),
  PoS (LATTICE2018) 311. 

\bibitem{B2D*:Nf2+1:Fermilab/MILC:wne1}
  A.~Bazavov {\it et al.} (Fermilab and MILC Collaboration),
  arXiv:2105.14019 [hep-lat].

\bibitem{B2D*:HMChPT:RW}
  L.~Randall and M.B.~Wise,
  Phys. Lett. B303 (1993) 135.

\bibitem{B2D*:HMChPT:S}
  M.J.~Savage,
  Phys. Rev. D65 (2002) 034014.

\bibitem{alphas:Nf2+1:JLQCD}
  K.~Nakayama, B.~Fahy and S.~Hashimoto,
  Phys. Rev. D94 (2016) 054507

\bibitem{luke_theorem}
  M.E.~Luke,
  Phys. Lett. B252 (1990) 447.

\bibitem{chiral_exp:Nf2:JLQCD}
  J.~Noaki {\it et al.} (JLQCD and TWQCD Collaborations),
  Phys. Rev. Lett. 101 (2008) 202004.

\bibitem{B2D*:Nf2+1:Fermilab/MILC:w1}
  J.A.~Bailey {\it et al.} (Fermilab and MILC Collaboration),
  Phys. Rev. D 89 (2014) 114504.

\bibitem{B2D*:HPQCD:w1}
  J.~Harrison {\it et al.} (HPQCD Collaboration),
  Phys. Rev. D 97 (2018) 054502.

\bibitem{shrinkage}
  C.~Stein, in Proceedings of the Third Berkeley Symposium
  on Mathematical Statistics and Probability, Vol. 1,
  pp. 197-206;
  O.~Ledoit and M.~Wolf, J.~Multivariate Anal. 88 (2004) 365.

\bibitem{CKM2021}
  T.~Kaneko,
  PoS (CKM2021) 048.


\bibitem{NPR}
  T.~Ishikawa, S.~Hashimoto and T.~Kaneko,
  PoS (LATTICE2021) 581 in these proceedings.

\bibitem{hT:eom}
  Y.~Sakaki, M.~Tanaka, A.~Tayduganov and R.~Watanabe,
  Phys. Rev. D 88 (2013) 094012.

\bibitem{hT:reso}
  D.~Melikhov and B.~Stech,
  Phys. Rev. D 62 (2000) 014006.
  
\end{thebibliography}
\end{document}